\documentclass[aps,pra,twocolumn,floatfix,amssymb]{revtex4-2}

\usepackage{graphicx,dcolumn,bm,xstring,braket,amsmath,comment,nicefrac,placeins}
\usepackage{color}
\usepackage[colorlinks,allcolors=blue,bookmarks=true]{hyperref}

\graphicspath{{./Figs/}} 

\newcommand{\const}{\mbox{const}}

\renewcommand{\bra}[1]{\left\langle #1 \right|}
\renewcommand{\ket}[1]{\left| #1 \right\rangle}
\renewcommand{\braket}[1]{\left\langle #1 \right\rangle }

\renewcommand{\Braket}[2]{\left\langle #1 \middle| #2 \right\rangle}

\newcommand{\beq}{\begin{eqnarray}}
\newcommand{\eeq}{\end{eqnarray}} 

\newcommand{\hide}[1]{}  %{{\textcolor{red}{[hide]}}}
\newcommand{\rmrkn}[1]{#1} %{{\textcolor[rgb]{0.6,0,0.1}{#1}}}
\newcommand{\rmrk}[1]{#1} %{{\textcolor[rgb]{0.6,0,0.1}{#1}}}
\newcommand{\TODO}[1]{} %{\bm{\textcolor[rgb]{0.0,0.6,0.2}{#1}}}
 
\newcommand{\Eq}[1]{{\textcolor{blue}{Eq.}}~\!\!(\ref{#1})} 
\newcommand{\Sec}[1]{{\textcolor{blue}{Sec.}}~(\ref{#1})} 
\newcommand{\App}[1]{{\textcolor{blue}{Appendix}}~\ref{#1}} 
\newcommand{\Fig}[1] {{\textcolor{blue}{Fig.}}~\!\!\ref{#1}}
\newcommand{\sect}[1]{{\bf #1.-- }}

\newcommand{\hrefl}[2]{\href{#2}{(#1)}}

\begin{document}

\title{Tweezer interferometry with NOON states}

\author{Yehoshua Winsten, Doron Cohen}
\email[Electronic address: ]{dcohen@bgu.ac.il}
\affiliation{Department of Physics, Ben-Gurion University of the Negev, Beer-Sheva 84105, Israel}

\author{Yoav Sagi} 
\email[Electronic address: ]{yoavsagi@technion.ac.il}
\affiliation{Physics Department and Solid State Institute, Technion - Israel Institute of Technology, Haifa 32000, Israel}

%\date{\today}

\begin{abstract}
Atomic interferometers measure phase differences along paths with exceptional precision. Tweezer interferometry represents a novel approach for this measurement by guiding particles along predefined trajectories. This study explores the feasibility of using condensed bosons in tweezer interferometry. Unlike the factor $\sqrt{N}$ enhancement expected with classical ensembles, using NOON state interferometry can yield an enhancement by a factor of $N$. We consider a protocol for a tweezer-based NOON state interferometer that includes adiabatic splitting and merging of condensed bosons, followed by adiabatic branching for phase encoding. Our theoretical analysis focuses on the conditions necessary to achieve adiabaticity and avoid spontaneous symmetry breaking. Additionally, we demonstrate the feasibility of the proposed scheme and estimate the time required to perform these sweep processes.   
\end{abstract}

\maketitle

%%%%%%%%%%%%%%%%%%%%%%%%%%%%%%%%%%%%%%%%%%%%%%%%%%%%%%%%
%%%%%%%%%%%%%%%%%%%%%%%%%%%%%%%%%%%%%%%%%%%%%%%%%%%%%%%%

%%%%%%%%%%%%%%%%%%%%%%%%%%%%%%%%%%%%%%%%%%%%%
\section{Introduction}

%Michelson-Morley experiment \cite{mme} 

%Gravitational-Wave Observatories (LIGO, Virgo and KARGA) - the most advanced interferometer ever built \cite{ligo} 

%interference has been demonstrated using a wide range of masses, including electrons, atoms, and complex molecules \cite{interferometry1, interferometry2} 

%To coherently split wave-functions of atoms, Atomic interferometers have initially employed diffraction from periodic fabricated structures \cite{AIF1, AIF2} and optical lattices \cite{AIF3}. 

%One axis twist dynamics\cite{OAT1,OAT2,OAT3} 

%NOON state phase \cite{NOON1,NOON2,NOON3} 

%splitting in BEC \cite{NonAdSplit,NonAdSplit2} 

%BJJ tunneling \cite{BJJtunneling1,BJJtunneling2} 

%Nonlinear atom interferometer (Ramsey) \cite{NonLinInterferometer}

Interferometry with matter waves \cite{RMV2018,interferometry1,interferometry2} is a sensitive method for measuring spatial variations in a force field or differential acceleration. This technique involves splitting an atomic wave packet \cite{AIF1,AIF2,AIF3,NonAdSplit,NonAdSplit2}, allowing it to propagate along two different paths, and then recombining it to estimate the relative phase shift acquired along the paths. The specific state of the wave function after splitting influences the interferometer's sensitivity. When the two parts of the wave packet are entangled, the state is referred to as non-classical \cite{NOON1,NOON2,NOON3}. Interferometry with non-classical states enables improvement in relative phase sensitivity, which can approach the Heisenberg limit $1/N$, where $N$ is the number of particles, \rmrk{see} \cite{noon-opt}. However, achieving this improvement in an experimentally feasible scenario is still an open challenge, \rmrk{mainly addressed so far in photonic systems \cite{opt1,opt2,opt3}.}

In the majority of atom interferometers, splitting, mirroring and recombining occur in momentum space through the absorption of photons, with the wave packets undergoing ballistic motion in between these processes \cite{interferometry1}. 
\rmrk{Recently}, a guided atomic interferometry scheme based on optical tweezers has been proposed \cite{tweezers1,tweezers2,tweezers3}. This method employs micro-optical traps to coherently split the wave function of atoms, guide them along selected trajectories, and then merge them to detect the interference pattern. The original proposal \cite{tweezers1} suggested preparing approximately ${N \sim 100}$ fermions simultaneously in different vibrational states of the tweezer, to then engage in the interferometric sequence in parallel. This approach offers a sensitivity enhancement by a factor of $\sqrt{N}$ compared to the single atom scenario. 

In this work, we explore the scenario of trapping bosons, specifically a Bose-Einstein Condensate (BEC), as opposed to fermions. This naturally introduces the possibility for novel methods for interferometry \cite{RMV2018,NonLinInterferometer,OAT1,OAT2,OAT3}.   
We focus in this work on NOON state interferomerty, where the bosons are in a symmetric superposition of all atoms being in a BEC in only one of the interferometer's paths.

We analyze a complete guided interferometry scheme using NOON states \cite{CL1}. To generate these states, one exploits the tunability of the interaction strength \cite{BJJtunneling1,BJJtunneling2} and the confining potential. In the presence of attractive interactions, an adiabatic change from a single well to a double well generates a NOON state between the two wells. After separating the wells, and letting the wave packet in each of them to acquire a distinct phase, the challenge lies in recombining them to realistically estimate the phase difference. By 'realistic' we mean that there should be no demand for accuracy in any experimental parameter that scales exponentially with $N$. Merely reversing the many-body splitting process does not meet this criterion. Instead, we incorporate a recombination method that employs a two-mode adiabatic following, which translates the relative phase into the probability of finding \emph{all atoms} at one of the interferometer's exit ports. Importantly, this scheme requires that the dynamic range of control over system parameters scales only linearly with~$N$. This aspect makes our proposal plausible for experimental realization.

\begin{figure*}
\includegraphics[width=\textwidth]{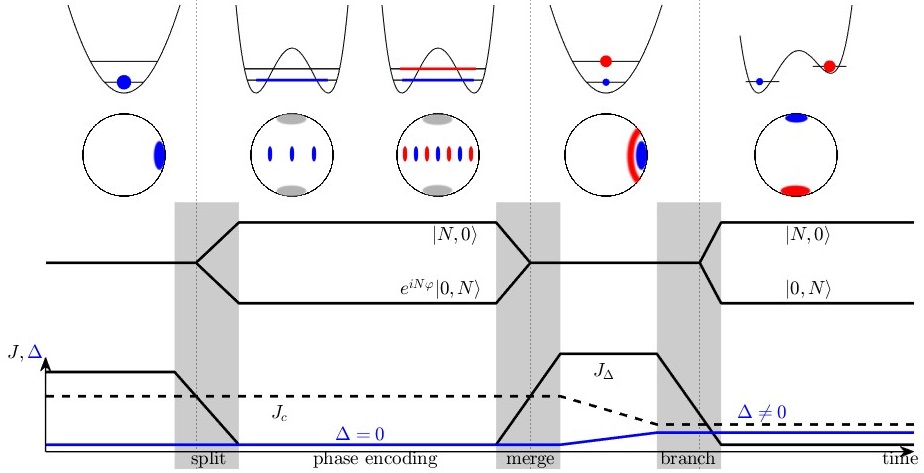} 
\caption{
{\bf The interferometry protocol.} 
The control parameters are $J(t)$ and $\Delta(t)$.  Their time dependence is plotted in the bottom panel (black and blue lines, respectively). The number of particles $N$ and the interaction $U$ are kept constant. The critical value of $J$ (see text) is indicated by dashed line. 
We start with all the particles condensed in the lower symmetric orbital of the dimer.  
The stages are:  
(a)~Decreasing~$J$ leads to adiabatic splitting; 
(b)~Phase encoding; 
(c)~Increasing~$J$ leads to partial or full merging; 
(d)~Increasing~$\Delta$ to a non-zero small value; 
(e)~Decreasing~$J$ again leads to adiabatic branching. 
In the proposed scheme (see text) $\dot{J}$ is adaptive.
The 3 upper rows of panels display: The potential landscape, where the vertical axis is energy. \rmrk{Note that the horizontal axis is $S_z$, the occupation imbalance, and not real space;} the Bloch sphere representation of the evolving state [sphere coordinates are $(S_x,S_y,S_z)$]; and the schematic bifurcation diagram. 
Further explanation of the figure, and of the Bloch sphere representation, is provided throughout the following sections.
}    
\label{fProtocol}  
\end{figure*}

{\bf Outline.--} 
In section~\ref{sec:TwoMode} we summarize the basics of two mode interferometry. We then introduce the protocol of \Fig{fProtocol} in Section~\ref{sec:Protocol}.  The Hamiltonian modeling of the system is presented in Section~\ref{sec:Hamiltonian}. Then in section~\ref{sec:Simulations} we provide results of simulations that motivate the further analysis and the suggested optimization procedure in Section~\ref{sec:Optimized} and Appendix~\ref{sec:Crossover}. The feasibility of the protocol and its comparison to to other proposed schemes are discussed in Sections~\ref{sec:Feasibility} and~\ref{sec:Discussion}.

%%%%%%%%%%%%%%%%%%%%%%%%%%%%%%%%%%%%%%

%%%%%%%%%%%%%%%%%%%%%%%%%%%%%%
\begin{figure}[!b]
\includegraphics[width=7cm]{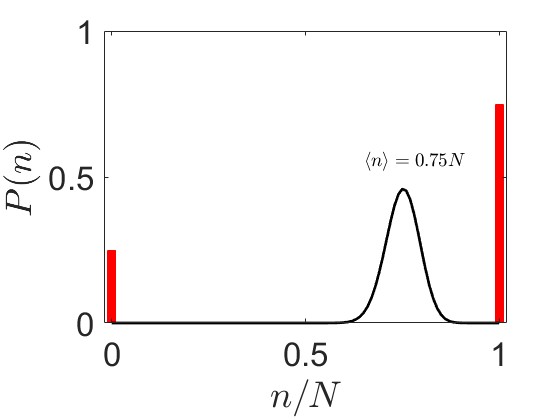} 
\caption{
{\bf Phase estimation.} 
Red line indicates the expected $P(n)$ of the suggested interferometry protocol with idealized NOON states. For the sake of comparison, the black curve is the probability distribution that one would observe in standard interferometry. The phase can be estimated from the average $\braket{n}$, or better, by fitting to the expected $P(n)$. With our protocol, the estimation requires multiple runs to asses the probabilities $P(0)$ and $P(N)$, \rmrk{see \Sec{sec:Protocol} for details}.     
}    
\label{fProb}  
\end{figure}

%%%%%%%%%%%%%%%%%%%%%%%%%%%%%%%%%%%%%%%%%%%%%
\section{Two-Mode Interferometry} 
\label{sec:TwoMode}

We define our objectiv as measuring a potential difference, $V$, between two regions (``sites"). If we have a single particle in a superposition of being in the two sites, the accumulated phase difference during time $\tau_{\text{enc}}$,  
the so-called {\em phase encoding} stage \cite{RMV2018}, 
is $\varphi=V\tau_{\text{enc}}$, where we employ units such that $\hbar=1$.  
When considering $N$ bosons in two sites, it is convenient to employ the basis states $|N{-}n,n\rangle$, indexed by the ground state occupation of the two sites. Any initial state can be written as a superposition $|\psi\rangle  = \sum_n c_n |N{-}n,n\rangle$. After phase encoding it evolves into  
\beq \label{ePrep}
| \psi(t_{\text{enc}}) \rangle \ \ = \ \ \sum_n c_n e^{i (n \varphi + \phi_n)} |N{-}n,n\rangle \, \, ,
\eeq
where $\phi_n$ is an additional phase that may arise due to inter-particle interactions. In the common scenario where the goal is to determine $\varphi$, it is better to have a design such that the interactions are zero ($\phi_n=0$) during the phase encoding stage. The best sensitivity is expected when preparing the system in a NOON state, namely, ${|\psi\rangle= \frac{1}{\sqrt{2}}\left[ |N,0\rangle  +  |0,N \rangle  \right]}$, for which 
\beq \label{ePrepC}
|\psi(t_{\text{enc}})\rangle \ \ =  \ \ \frac{1}{\sqrt{2}}\left[ |N,0\rangle  +  e^{i N \varphi} |0,N \rangle  \right] \, \, .
\eeq
Here, we omitted the $\phi_n$, because even in the presence of interaction ${\phi_N=\phi_0}$, assuming that the interaction is the same in both sites.  In spite of its obvious benefits, NOON interferometry requires overcoming two challenges: preparation of the initial state, and performing the {\em phase estimation}.

It is convenient to regard that basis states   
${ |N{-}n,n\rangle }$ with ${n=0, \cdots, N}$ 
as polarization states of a spin entity 
with ${S_z=+(N/2),\cdots,-(N/2)}$ respectively. 
The set of all (full) polarization direction 
is conveniently represented on the Bloch sphere.
Then we define the following states of interest, 
and name them accordingly: 
\beq
\rmrk{\ket{Z}} &=& \text{NorthPole}  \equiv |N,0\rangle \ = \ \frac{1}{\sqrt{N!}}(a_1^{\dag})^N |0,0\rangle \\ 
\rmrk{\ket{\bar{Z}}} &=& \text{SouthPole}  \equiv |0,N\rangle \ = \ \frac{1}{\sqrt{N!}}(a_2^{\dag})^N |0,0\rangle  
\\ \label{eXstate}
\rmrk{\ket{X}}  &\equiv& \frac{1}{\sqrt{N!2^N}}(a_1^{\dag} + a_2^{\dag})^N |0,0\rangle \\
\rmrk{\ket{\bar{X}}}  &\equiv& \frac{1}{\sqrt{N!2^N}}(a_1^{\dag} - a_2^{\dag})^N |0,0\rangle \\
\rmrk{\ket{+}} &=& \text{EvenCat}  \equiv \frac{1}{\sqrt{2}}\left[ |N,0\rangle  +  |0,N \rangle \right] \\
\rmrk{\ket{-}} &=& \text{OddCat}  \equiv \frac{1}{\sqrt{2}}\left[ |N,0\rangle  -  |0,N \rangle \right]
\eeq
It is important to realize that a $\pi/2$ rotation of $\ket{Z}$ state,  which is generated by $S_y$, yields an $\ket{X}$ coherent state and not a $\ket{+}$ cat state. The preparation of the latter can be performed using a non-linear splitter, as we explain below.

An interferometry protocol includes four steps: Preparation of the state $\psi(0)$; Phase encoding stage ${\varphi=V\tau_{\text{enc}}}$; Transformation of the state $\psi(t_{\text{enc}})$ by the atomic beam combiner; Measurement of the number of atoms in each of the two output ports. By repeating this procedure many time, one can construct the probability distribution, $P(n)$, of the number of atom in one of the ports (it does not matter which). The objective of the interferometry protocol is to deduce $\varphi$ from $P(n)$. 

In standard interferometry, one starts with a coherent state that is represented by a Gaussian-like distribution at the North pole of the Bloch sphere. A Ramsey $\pi/2$ pulse rotates it to $\ket{X}$, where ${\varphi=0}$. During the time~$t$ it rotates along the Equator to ${\varphi=Vt}$. A second Ramsey $\pi/2$ pulse follows. 
The obtained $P(n)$ is illustrated in \Fig{fProb}, where the expectation value of $n$ is related to $\varphi$, namely,   
\beq
\braket{n} \ \equiv \ \sum_n n P(n)  
\ = \ [1-\cos(\varphi)]\frac{N}{2} \ \ .
\eeq
A single run of the experiment provides an estimate for~$\braket{n}$, and hence for~$\varphi$, with relative uncertainty that scales like~$1/\sqrt{N}$. This resolution reflects the width of the Gaussian-like distribution. If statistics from $M$ measurements is accumulated, the uncertainty reduces to $1/\sqrt{MN}$.

If one uses the same protocol, but replaces the coherent state by a NOON state, one gets a $P(n)$ whose line-shape is insensitive to $\varphi$. A more sophisticated protocol is required for both the NOON state preparation and for the phase estimation. We describe such protocol in the next section, and compare it to different alternatives in the concluding section.

%%%%%%%%%%%%%%%%%%%%%%%%%%%%%%%%%%%%%%%%%%%%%
\section{The protocol}
\label{sec:Protocol}

A way to prepare an approximate NOON state has been suggested in Ref. \cite{NonAdSplit}, and we shall refer to it as {\em non-adiabatic splitting}. The idea is to start with an $\ket{X}$ coherent state, and allow it to stretch dynamically such that after a certain time $P(n)$ is peaked at ${n{=}0}$ and ${n{=}N}$, and relatively negligible for other values of~$n$. 
\rmrk{The advantage of such protocol is that the evolution is generated by a time-independent Hamiltonian: it is a rapid quench process. The disadvantage is the need to have a precise control over the time duration of the quench, and the relatively low fidelity of the generated NOON state. The latter point is further discussed in the concluding section.}

The optional procedure for preparing a NOON state that we consider below is {\em adiabatic splitting}, as in \cite{CL1}. Unlike the non-adiabatic quench, it is not sensitive to timing issues. Furthermore, we shall see that it provides a higher fidelity NOON state, for which $P(n)$ is negligible for $n$ that is neither $0$ nor $N$.  
For the phase estimation we use the biased version of the same process, which we call {\em adiabatic branching}.

In this section, we briefly summarize the interferometry protocol, depicted in \Fig{fProtocol}. In the following sections, we will present a thorough analysis of it. The on-site interaction is denoted as $U$, with time units chosen such that $NU=1$. The time-dependent control parameters are the hopping (or tunneling) frequency between the two sites, $J(t)$, and the energy bias between the two sites, $\Delta(t)$.  \rmrk{We index the adiabatic levels $E_{\nu}$ such that the ground state is $\nu{=}1$.}

%%%%%%%%%%%%%%%%%%%%%%%%%%%%
\subsection{Stage (a) - adiabatic splitting}

\rmrk{The protocol begins with the atoms in the $\ket{X}$ state as defined in \Eq{eXstate}.} This is a coherent state, where all the bosons are condensed in the lowest orbital of the symmetric dimer. 
\rmrk{In the Bloch-Sphere language that we further elaborate in the next section, this state is represented by a Gaussian-like Hussimi function that is located at the bottom of a single-well potential. This potential (on the Bloch Sphere) should not be confused with the actual potential in space. Namely, we assume that the atoms are trapped at any time by a potential created by two optical tweezers that generate two potential minima in real space. In this work we model only the case of two separate minima. The control over the tunnel coupling $J$ can be achieved by changing the distance between the minima or the depth of each of the potentials.}    

The goal of stage (a), during which $\Delta=0$, is to adiabatically transform this coherent state into an EvenCat superposition $\ket{+}$. This is the process that we call {\em adiabatic splitting}.
\rmrk{In \rmrkn{section}~\ref{sec:Hamiltonian} we explain that this splitting is achieved because as $J$ is decreased the single well potential in the Bloch sphere bifurcates and becomes a double-well. This process is known as self-trapping. What we get is a superposition of self-trapped particles.} 

%%%%%%%%%%%%%%%%%%%%%%%%%%%%
\subsection{Stage (b) - phase encoding}

\rmrk{During the phase encoding stage the external bias leads to the accumulation of phase difference $N\varphi$ between the two sites.} Thus, at the end of the phase encoding stage the state of the system is 
\beq \label{ePhiState}
\ket{\varphi} \ \ = \ \ \frac{1}{\sqrt{2}} 
\left[\ket{Z}  + e^{iN\varphi} \ket{\bar{Z}}\right]
\eeq
This is merely re-writing of \Eq{ePrepC}.
The Even and Odd cat states $\ket{\pm}$ are special cases of \Eq{ePhiState}, that correspond to ${N\varphi=0,\pi}$.

The phase encoding stage, by itself, is formally a non-adiabatic process that induces Rabi oscillation between the even and odd quasi-degenerate cat states. These states are indexed ${\nu=1,2}$ respectively. Ideally, higher adiabatic levels (${\nu>2}$) are not involved.   
\rmrk{It is implicitly assumed that the bias is not strong enough to mix these quasi-degenerate states with the higher levels, \rmrkn{nor does it have a strong enough gradient to scramble the orbitals within each location.} This ensures that a two orbital approximation is adequate. Note that the tunnel-splitting of the $\ket{\pm}$ states is exponentially small in $N$ and therefore negligible compared to the spacing to higher levels.}

%%%%%%%%%%%%%%%%%%%%%%%%%%%%%%%%%%%%%%%%%%
\subsection{Stage (c) - merging}

During the merging stage $J$ is increased, and the double well on the Bloch Sphere becomes again a single well (no self-trapping). \rmrk{In the absence of phase accumulation the adiabatic evolution leads back to the ground state $\ket{X}$, meaning that all the $N$ particles are condensed in the lower orbital. If we have e.g. phase accumulation ${N\varphi=\pi}$, it means that the system has evolved into the second adiabatic state ${\nu=2}$, and therefore after merging, we have $N{-}1$ particles in the lower orbital, and one particle in the excited orbital.} More generally, due to phase accumulation the system ends up in a superposition of the first and second adiabatic levels. The first adiabatic level ($\nu=1$) is the ground state (condensation in the lowest orbital), while the second adiabatic level ($\nu=2$) is a one-particle excitation.

%%%%%%%%%%%%%%%%%%%%%%%%%%%%%%%%%%%%%%%%%%
\subsection{Stage (d) - Setting bias}

\rmrk{Stage (d) looks rather innocent. A very small bias is applied, meaning that there is a potential difference $\Delta$ between the two sites.  Using Bloch Sphere terminology it means that the single well minimum is no longer at ${S_x=N/2}$, but slightly shifted.  This small bias has no significant effect as long as $J$ is kept large. It becomes important only as $J$ is decreased.}

%%%%%%%%%%%%%%%%%%%%%%%%%%%%%%%%%%%%%%%%%%
\subsection{Stage (e) - adiabatic branching}

The branching stage is formally similar to the splitting stage, but here the bias is non-zero ($\Delta \ne 0$).  If, at the end of stage (c), the state is an even state ($\nu=1$), it is mapped after stage (e) to $|N,0\rangle$, whereas if, due to phase accumulation, the state at the end of stage (c) becomes an odd state ($\nu=2$), it is mapped to $|0,N\rangle$.

More generally, for any given $\varphi$, this protocol yields a superposition of the $|N,0\rangle$ and $|0,N\rangle$ states, with the probabilities reflecting the relative phase. Therefore, $P(n)$ has peaks at $n=0$ and at $n=N$, see \Fig{fProb}.

%%%%%%%%%%%%%%%%%%%%%%%%%%%%%%%%%%%%%%%%%%
\subsection{\rmrk{Phase estimation}}  

To estimate $\varphi$, a single measurement is insufficient as it does not provide $P(n)$. One needs to perform $M \gg 1$ measurements to determine the relative height \rmrk{$P(N)/P(0)$} of the two peaks in \Fig{fProb}. 
\rmrk{This does not contradict the well known statement \cite{noon-opt} that NOON state interferometry achieves the Heisenberg limit. The sensitivity scales as $1/N$ because the accumulated phase in the phase encoding stage is $N\varphi$. But in some sense there is also an $M$ dependent prefactor that determines the accuracy.}

An important property of our protocol is that at each run all atoms should be found at one and only one of the output ports. Any deviation from that rule signals a problem in the data point, which can then be discarded. Thus, the protocol implies a natural way to improve the quality of the results by post-selection.

%%%%%%%%%%%%%%%%%%%%%%%%%%%%%%%%%%%%%%%%%
\begin{figure}
\centering
\includegraphics[width=3.5cm]{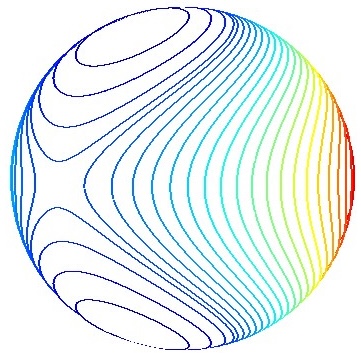}
\includegraphics[width=8.5cm]{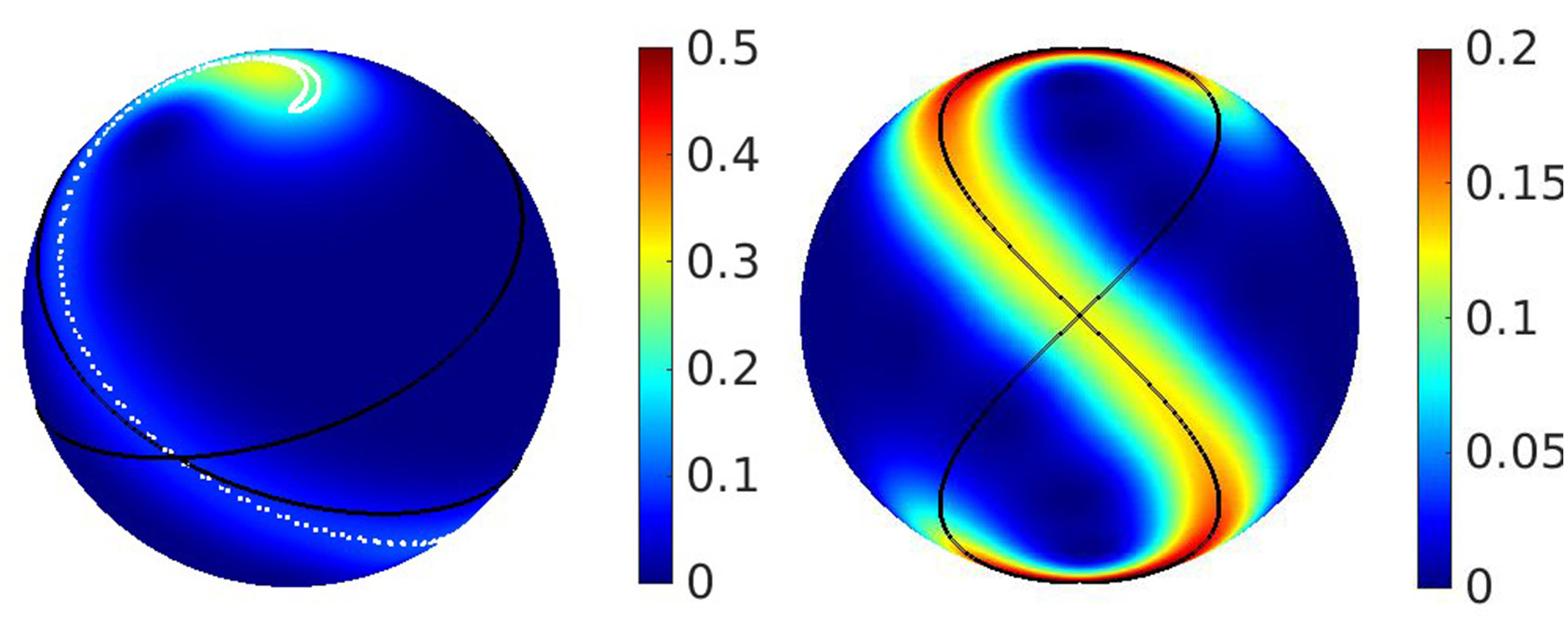}
\caption{
{\bf Energy landscape and wavepacket dynamics.} 
 The phasespace of the dimer system is the two-dimensional Bloch sphere whose embedding coordinates are ${(S_x,S_y,S_z)}$.
 {\bf Upper panel:} WKB energy contours ${\mathcal{H}_{\text{dimer}}=E_{\nu}}$ are plotted on the Bloch sphere. 
 We use units of time such that $NU{=}1$. The coupling $J{=}0.4$ is larger than $J_c$, while $\Delta{=}0$ and $N=30$.
{\bf Lower panels:} 
\rmrkn{Husimi functions -- probability to find the system in an ${(S_x,S_y,S_z)}$ coherent state is indicated by color.} 
{\bf Right panel:}
The Husimi function of an approximate cat state that is produced using a non-adiabatic splitting protocol.
The initial state (not displayed) is a condensate $\ket{X}$ with ${N=30}$ particles, located at an unstable (hyperbolic) point of the Bloch sphere.
We set ${J=0.5}$, hence the non-linearity (${u=2}$) is such that the separatrix (black line) extends up to the North and South Poles. The time of the evolution is $\pi/\omega_x$, to maximize the splitting.
{\bf Left panel:} 
Snapshot of the evolving Husimi function
during a non-optimal adiabatic splitting of the same condensate at ${J(t)=0.1}$. 
The sweep rate is ${\dot{J} = 1/8}$.
The Bloch sphere is slightly rotated  
such that the North pole is fully visible 
(the hidden South pole supports the other half of the distribution).   
The white points represent a cloud that is evolved 
using classical equations of motion.
The initial classical cloud is a circle 
of radius ${n_0 = 1-(4N)^{-1}}$. 
The outcome of this simulation when ${J(t)=0}$ 
is approximately a NOON state with probability ${P(n) \approx 0.23}$
for getting either $n{=}0$ or $n{=}N$. 
For an ideal protocol one aims at getting ${P(N) \approx P(0) \approx 0.5}$.
% The classical simulation implies ${ \theta \sim 0.46 \theta_{adiabatic} }$.  
}  
\label{fig:husimi}  
\end{figure}

%%%%%%%%%%%%%%%%%%%%%%%%%%%%%%%%%%%%%%%%%%%%%%%%%%%%%%%%%%%%%%%%%%%%%%%%%%
\section{Hamiltonian formulation}
\label{sec:Hamiltonian}

The dimer Hamiltonian can be written using generators of spin-rotations. One starts with the standard Bose-Hubbard version, namely,   
\beq
\mathcal{H}_{\text{dimer}} = && \sum_{j=1,2} \left[ \epsilon_j a^{\dag}_j a_j 
- \frac{U}{2} a^{\dag}_ja^{\dag}_ja_ja_j  \right] 
\\
&& -\frac{J}{2} \left( a^{\dag}_2a_1 + a^{\dag}_1a_2 \right)
\eeq
where $a_j$ and $a_j^{\dag}$ are the annihilation and creation operators of site ${j=1,2}$. The observable $S_z$ is defined as half the occupation difference in the site representation, 
namely 
%there is a mistake here
${S_z=\frac{1}{2}[a^{\dag}_1 a_1-a^{\dag}_2 a_2]}$. 
Then, one defines ${S_{+}=a^{\dag}_1 a_2}$ and ${S_{-} = S_{+}^{\dag}}$, 
and the associated $S_x$ and $S_y$ operators.
Accordingly, $S_x$ is identified as half the occupation difference in the momentum representation, where the momentum states are the mirror-symmetric even and odd orbitals.

With the above notations 
the dimer Hamiltonian takes 
the following form, 
\beq  \label{eHdimer}
\mathcal{H}_{\text{dimer}} 
=
- U S_z^2 
- \Delta S_z
- J S_x \, \, ,
\eeq
where ${\Delta=\epsilon_2-\epsilon_1}$ is the energy bias between the sites. 
For the subsequent analysis we define
\beq
n &\equiv& \frac{N}{2}-S_z \\
n_{\text{ex}} &\equiv& \frac{N}{2}-|S_z|
\eeq
The value $n_{\text{ex}}=0$ indicates that the state of the system is fully contained in the subspace 
that is spanned by the ${|N,0\rangle}$ and ${|0,N\rangle}$ states. We consider below adiabatic processes whose outcome is supposed to be in this subspace. Accordingly non-zero $\braket{n_{\text{ex}}}$ serves as a measure for non-adiabaticity.

In a quantum context $(S_x,S_y,S_z)$ are operators \rmrk{of an $S=N/2$ spin entity}, while in a semiclassical perspective these are ``phasespace" coordinates of the so-called Bloch Sphere, with \rmrk{${S_x^2+S_y^2+S_z^2 = (S{+}1)S \approx (N/2)^2 }$}. The contour lines ${ H(S_x,S_y,S_z) = E }$ are the skeleton on which the quantum states are constructed. In a WKB picture (see \App{appA}) each contour line supports an eigenstate, and the area spacing between contour lines equals the Planck cell area $h \approx 2\pi$. See the upper panel of \Fig{fig:husimi} for a demonstration.  Any state of the system can be represented on the Bloch sphere using e.g. Wigner or Husimi representation.  

Without loss of generality, we assume an attractive interaction. In our sign convention it means ${U>0}$.
A dimensionless parameter that determines the underlying phase-space structure is 
\beq
u \ \ = \ \ \frac{NU}{J} \, \, .
\eeq
For $u{=}0$ the energy landscape \rmrk{on the Bloch sphere} has one minimum (at ${S_x=N/2}$),  and one maximum (at ${S_x=-N/2}$).  
Considering an unbiased system with $\Delta=0$, as $J$ is decreased, a bifurcation appears at ${S_x=N/2}$ once $u$ becomes larger than unity. At this point, \rmrk{a double well structure is formed}, as illustrated in the upper panel of \Fig{fig:husimi}.  The minima are located at $S_x = u^{-1}(N/2)$ with $S_z=\pm \sqrt{1-u^{-2}}(N/2)$, while $S_y=0$. In particular, for $J=0$ the minima are at the North and South poles.

For non-zero $\Delta$ the bifurcation happens at  
\beq \label{eEcDia} 
J_c = \left[ (NU)^{2/3} - \Delta^{2/3} \right]^{3/2} \ \ .
\eeq   
For $J \lesssim J_c$ the potential becomes \rmrk{an asymmetric double-well structure} with a secondary minimum that is born at the vicinity of the ${S_x=N/2}$. The bifurcation for ${\Delta\ne 0}$ is of a saddle-node type, instead of the ${\Delta=0}$ pitchfork bifurcation that generates a symmetric double well.

%%%%%%%%%%%%%%%%%%%%%%%%%%%%
\begin{figure}
\includegraphics[width=7cm]{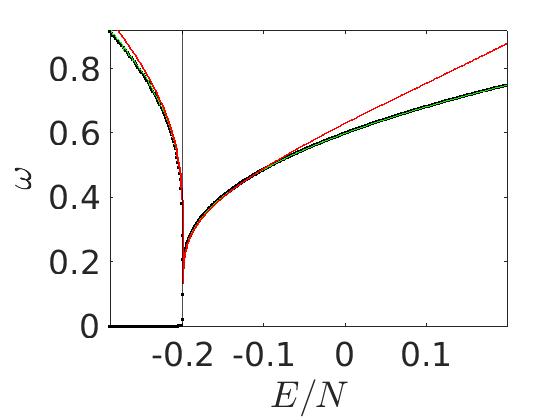}
\includegraphics[width=7cm]{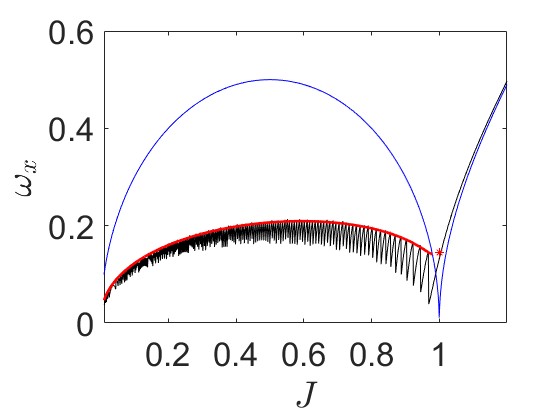}
\caption{
{\bf Dimer frequencies:} 
Upper panel: The dependence of $\omega(E)$ on the energy is extracted (black dots) from the difference $E_{\nu{+}1}-E_{\nu}$ for $N=500$ particles. \rmrkn{The dimensionless energy and frequency correspond to choice of units such that ${NU=1}$, while $J=0.4$.} Here $\Delta=0$ and therefore the lower part of the spectrum consists of doublets, which is reflected by the appearance of an $\omega \sim 0$ branch. The exact and approximated analytical results \Eq{eq:dA} and \Eq{eq:omegaE} are represented by green and red lines.    
Lower panel: The characteristic frequency near the separatrix ($\omega_x$) is plotted as a function of $J$. The Black line is from the calculation of energy differences; the blue line is $\omega_J$ of \Eq{eq:omegaJ}; and the red line is the analytical estimate \Eq{eq:omegaX}. The red asterisk is \Eq{eq:omegaC}.
}  
\label{f:Sz2-Sx}  
\end{figure}

The stationary states of the Hamiltonian have energies $E_{\nu}$. A superposition of two consecutive non-degenerate eigenstates oscillates with a frequency $\omega(E)$ that is determined 
by the level spacing $E_{\nu{+}1}-E_{\nu}$. \Fig{f:Sz2-Sx} demonstrates the energy dependence of $\omega(E)$, which can be analytically determined   
using a WKB procedure, see \App{appA}. 
The characteristic frequency ${\omega_0 \equiv NU }$ 
serves to fix the time units in our simulations. 
For ${J>J_c}$ and ${\Delta=0}$ the frequency at the minimum energy is 
\beq \label{eq:omegaJ}
\omega_J \ = \ \sqrt{|J-NU|J} \ = \omega_0  \sqrt{\left|\frac{1}{u}-1\right|\frac{1}{u}} \, \, .
\eeq
For $J=J_c$, it vanishes in the large $N$ limit, while the WKB approximation provides the scaling 
\beq \label{eq:omegaC}
\omega_c \ \propto  \frac{1}{N^{1/3}} \, \omega_0 \, \, .
\eeq  
For $J>J_c$ the frequency $\omega_J$  
is the instability exponent of the hyperbolic point.
The actual frequency along the separatrix 
that originates from this point 
can be estimated  (see \App{appA})
\beq \label{eq:omegaX}
\omega_x \ = \ \left[ \frac{1}{\pi} \ln \left(\frac{16 \omega_J^3}{NJU^2}\right) \right]^{-1} \omega_J \,  \, .
\eeq
Note that for large $u$ the argument 
of the log function becomes ${\propto \, N/\sqrt{u}}$ in agreement with the pendulum approximation of Ref.~\cite{csd}. 

Thus, the frequency $\omega_x$ of \Eq{eq:omegaX} becomes {\em zero} in the classical limit (${N=\infty}$), but away from the bifurcation it is finite and of order $\omega_J$ for any realistic finite~$N$.  This observation is crucial for the design of dynamical splitting. In the non-adiabatic scheme, the time to get an approximate cat state is $\pi/\omega_x$. The outcome is illustrated on the right side of the lower panel of \Fig{fig:husimi}. On the left side of the same panel we display the outcome of an {\em adiabatic} splitting process. In the latter case, a finite level spacing is an essential condition.  The bottleneck of the adiabatic splitting process is during the bifurcation at ${S_x=N/2}$, which happens when the separatrix is born,  where \Eq{eq:omegaC} rather than \Eq{eq:omegaX} determines the level spacing. We discuss and analyze the implied adiabatic condition in Section~\ref{sec:Optimized}.

%%%%%%%%%%%%%%%%%%%%%%%
\begin{figure}[!h]
\includegraphics[width=7cm]{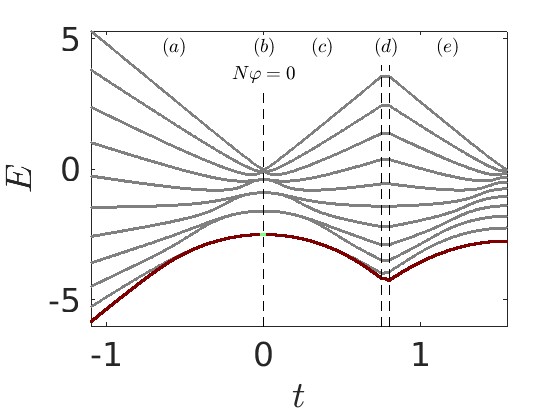} 
\includegraphics[width=7cm]{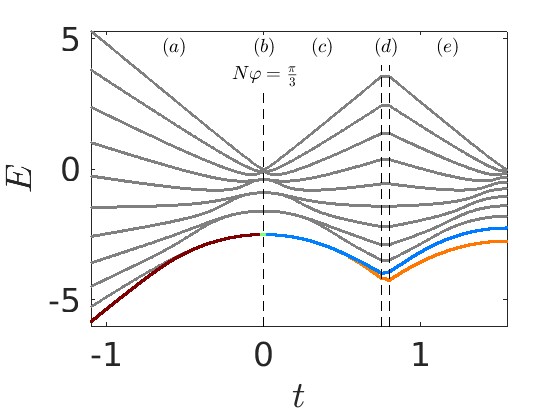} 
\includegraphics[width=7cm]{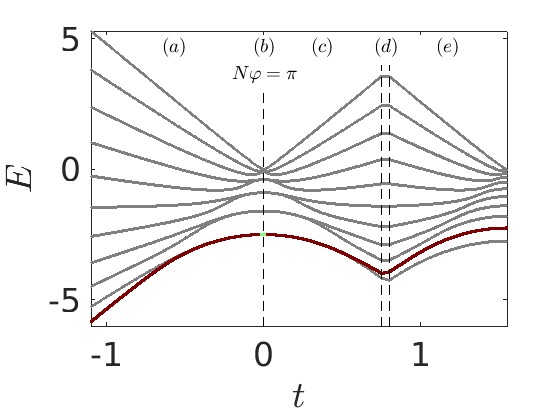} 
\caption{
{\bf Adiabatic evolution.}
Illustration of representative simulations with $N{=}10$ particles. \rmrk{The stages (a)-(e) of the protocol are indicated. The adiabatic levels $E_{\nu}$ are plotted as a function of time.} The evolving state is projected on the adiabatic states. The probabilities are color-coded.  
If the probability $p_{\nu}$ is negligibly small, the color is gray. Red color indicates ${p\approx 1}$.  
The accumulated (added) relative phase at $t{=}0$ 
is $\varphi{=}0$ (upper) 
and $N\varphi{=}\pi/3$ (middle) 
and $N\varphi{=}\pi$ (lower).   
\rmrkn{The units of time and energy are chosen such that ${NU=1}$.}   
The sweep rate is $\dot{J}=10^{-4}$ and $\dot{\Delta}=10^{-4}$, 
while $J(\text{max}){=}1.1$, and $\Delta(\text{max})=U/2$.
}  
\label{fig:EvsTime}  
\end{figure}

%%%%%%%%%%%%%%%%%%%%%%%%%%%%%
\begin{figure}[!h]
\includegraphics[width=6cm]{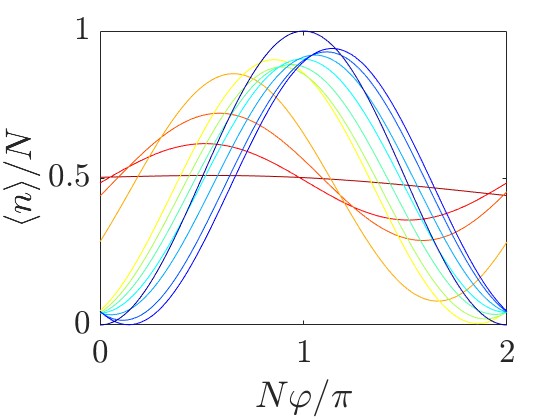}

\includegraphics[width=4cm]{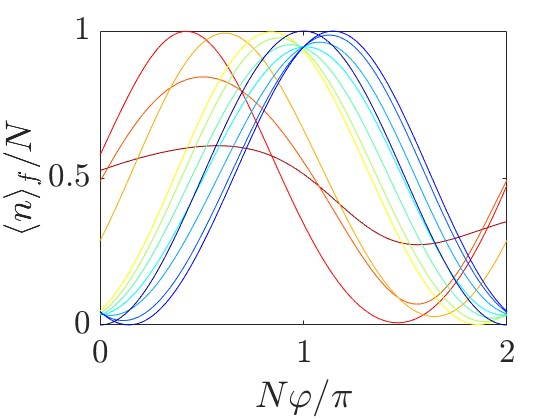} 
\includegraphics[width=4cm]{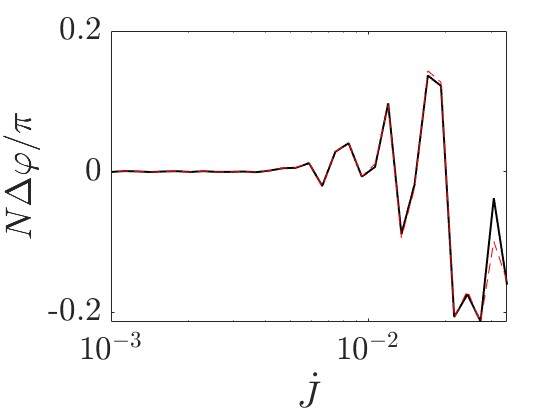} 

\caption{
{\bf Dependence of $\braket{n}$ in the suggested protocol on the encoded phase, $\varphi$.} 
In the upper and in the left lower panels, respectively, the normalized average occupation $\braket{n}/N$ is calculated either without or with rejection of the ``bad" runs, see \Eq{eEST}. The comparison shows improvement in the strength of the signal. 
The number of particles is ${N=30}$.
The curves are for different sweep rates. 
We use the same constant $\dot{J}$ for splitting, merging and branching. 
From red to blue ${\dot{J}=1/5,1/7,1/10,1/20,1/61,\cdots,1/67,1/600}$. 
In the right lower panel we display the $\dot{J}$ dependence of the systematic error in the phase estimation (dashed line, barely resolved, is after rejection of ``bad" runs).
}  
\label{fig:nvsvarphi}  
\end{figure}

%%%%%%%%%%%%%%%%%%%%%%%%%%%%%%%%%%%%%%%%%%%%%
\begin{figure}
\includegraphics[width=6cm]{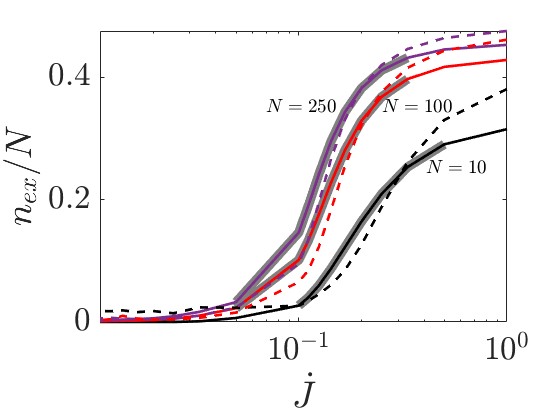}
\includegraphics[width=6cm]{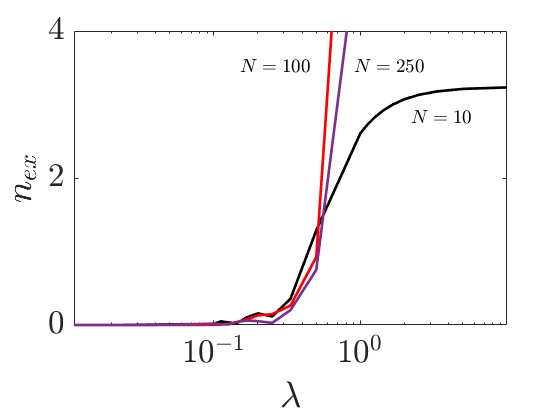}
\caption{
{\bf Non-adiabatic leakage to higher levels.}
The expectation value $\braket{n_{\text{ex}}}$ reflects 
the non-adiabatic leakage to higher energy levels (${\nu > 2}$).
Upper panel: $\braket{n_{\text{ex}}}$ is extracted 
from the Quantum simulations (solid lines),
for $N=10$ (black) and for ${N=100, 250}$.  
and compared with results 
of semi-classical simulations (dash lines).
The adiabatic-diabatic crossover, 
based on \Eq{eq:cdotJ},  
is indicated by gray background.    
Lower panel: Quantum results for an optimized split process,
with the same $N$, but with the adaptive sweep rate of \Eq{eq:dotJ}. 
Here the horizontal axis is the dimensionless sweep rate~$\lambda$, 
while the vertical axis is not scaled.  
}  
\label{fOutcome}  
\end{figure}

%%%%%%%%%%%%%%%%%%%%%%%%%%%%%%%%%%%%%%%%%%%%%%%%%%%%%%%%%%%%%%%%%%%%%%%%%%
\section{Simulations}
\label{sec:Simulations}

Simulations of the protocol requires to generate the evolution from a time dependent Hamiltonian $H(t)$. In the quantum simulations we simply evolve the initial state $\Psi(0)$ in very small time steps with the unitary operators $\exp[-i dt H(t)]$. In the semiclassical simulations we use the so-called  truncated Wigner approximation. Namely, the initial coherent state is represented by a Gaussian cloud of points on the Bloch Sphere, and then propagated using classical equations of motion. Mainly we are interested in the expectation value $\braket{n}$ at the end of the protocol.   

An illustration of representative simulations for the suggested protocol of \Fig{fProtocol} 
is presented in \Fig{fig:EvsTime}. In the adiabatic limit only the two lowest levels (${\nu=1,2}$) participate (as further explained in the next paragraph). 
The outcome depends on $\varphi$: 
If ${\varphi=0}$ the system stays in the ground state (${\nu=1}$), 
while if ${N\varphi=\pi}$, the state evolves to the next level (${\nu=2}$). 

To avoid confusion, it is important to realize that the phase accumulation stage is {\em not} regarded, and not simulated, as part of the adiabatic protocol. It acts as an instantaneous Rabi rotation with the same effect as that of the finite potential difference experienced by the two sites during the phase encoding stage. Thus, the phase accumulation can lead to a transition from an EvenCat state to an excited OddCat state. Nevertheless, this ``excitation" practically costs no energy, because the energy splitting between $E_1$ and $E_2$ during the phase encoding is exponentially small in $N$, due to the potential barrier.   

On the quantitative side, we calculate the probabilities 
\beq
p_{\nu}(t) \ \ = \ \ \left| \Braket{E_{\nu}(t)}{\Psi(t)} \right|^2 
\eeq 
where $\ket{E_{\nu}(t)}$ are the adiabatic levels, namely, the instantaneous eigen-energies  of the time dependent Hamiltonian $H(t)$. 
At the end of the protocol $J{=}0$ while ${\Delta\ne0}$, and accordingly the eigenstates of the non-degenerated Hamiltonian $H(t)$ are the occupation states ${|N-n,n\rangle}$. It follows that at the end of the simulation the probabilities $p_{\nu}$, up to indexing, are the probabilities $P(n)$ that we want to calculate.

{\bf Idealized protocol.--} 
As explained in the previous paragraph, in an ideal adiabatic protocol each run of the experiment ends with either ${n=0}$ or ${n=N}$.
In practice, due to lack of strict adiabaticity, or due to other disturbances we have non-zero probability to get other values of $n$.
\rmrk{This bring the idea that a rejection of bad measurements can improve the results. We discuss this option below. On top there might be errors that might not be as easily identified. We discuss briefly this point below as well.  }

{\bf Handling of explicit errors.--} 
\rmrk{By explicit error we mean getting in some runs illegal values of~$n$.} In \Fig{fig:nvsvarphi}
we show the dependence of $\braket{n}$ on $\varphi$  for representative sweep rates without or with rejection of the ``bad" runs. The ``bad" runs are those for which the outcome is neither ${n=0}$ nor ${n=N}$, indicating an undesired non-adiabatic transition. Due to this rejection we get a filtered estimate $\braket{n}_f$ instead of  $\braket{n}$, namely,   
\beq \label{eEST}
\braket{n} \ \ \mapsto \ \ \braket{n}_f = \frac{P(N)}{P(0)+P(N)} N 
\eeq
This leads to some improvement in the strength of the signal, though it does not overcome the systematic error due to non-adiabaticity, see right lower panel in \Fig{fig:nvsvarphi}.

{\bf Handling other systematic errors.--}
\rmrk{There are errors whose testing requires to go beyond the framework of the Bose-Hubbard modeling.} Let us focus on one example. Let us assume that there is dephasing during the phase-accumulation stage. \rmrk{Such dephasing can occur e.g. due to the loss of even a single particle.} Thus, instead of getting a coherent cat superposition ${\ket{\varphi}}$,  we get a mixture of ${ \ket{Z} \propto \ket{+} + \ket{-} }$ and ${ \ket{\bar{Z}} \propto \ket{+} - \ket{-} }$. The outcome of the latter cannot be distinguished from the outcome of ${\ket{\varphi=\pm \pi/2} \propto \ket{+} \pm i \ket{-} }$ because in all these cases we get equal probabilities for $n{=}0$ and $n{=}N$.  Accordingly, what we measure in practice is    
\beq
\frac{P(N)}{P(0)}   \ \ \mapsto \ \ \frac{(1-p)P(N)+p}{(1-p)P(0)+p}
\eeq
where $p$ is the probability to have a dephasing events (or loss of particle) during phase accumulation. The good news here is that such error has a systematic effect on the statistics, and therefore, after extraction of $p$, can be compensated. 

%%%%%%%%%%%%%%%%%%%%%%%%%%%%%%%%%%%%%%%%%%%%%
\begin{figure*}
\includegraphics[width=5cm]{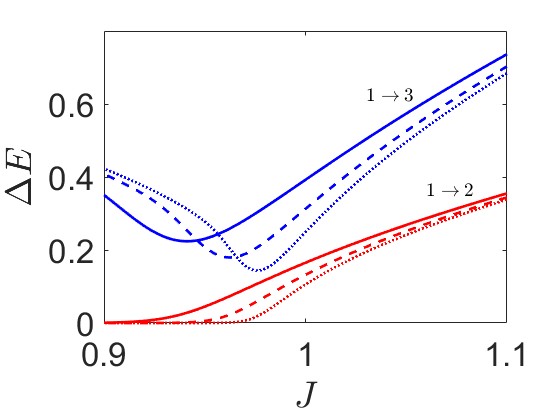} 
\includegraphics[width=5cm]{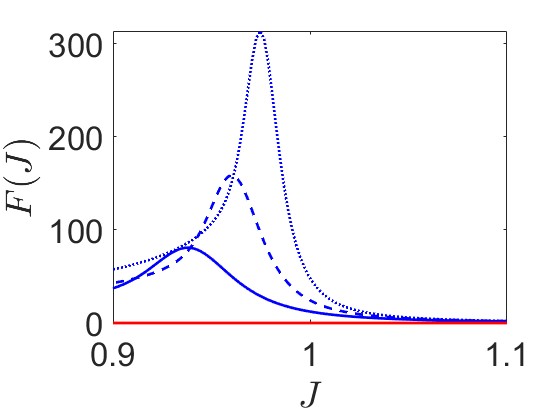} 
\includegraphics[width=5cm]{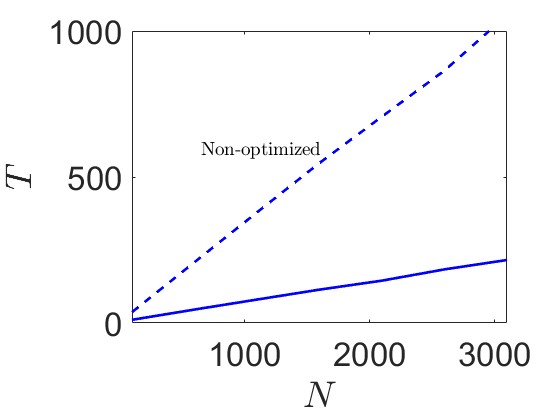} 
\\
\includegraphics[width=5cm]{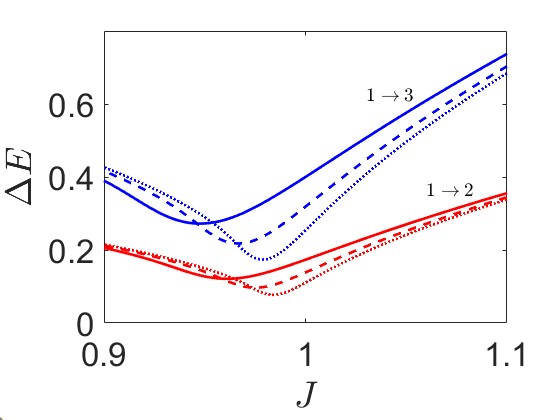} 
\includegraphics[width=5cm]{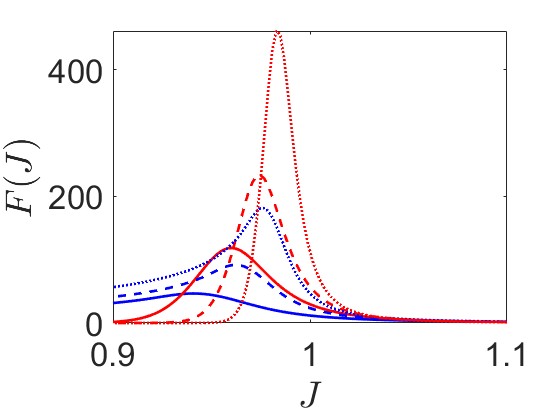} 
\includegraphics[width=5cm]{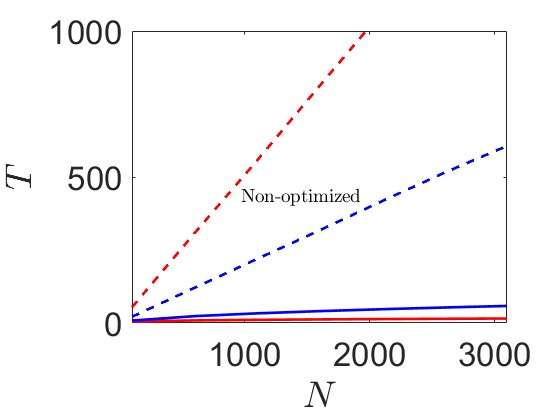}
\caption{
{\bf The adiabatic condition.} 
Upper row is for splitting process ($\Delta=0$), 
while lower row is for branching process ($\Delta \neq 0$).   
Blue and red lines concern transitions 
to $\nu{=}3$ and to $\nu{=}2$, respectively.   
Left panels: The energy difference ${\Delta E = E_{\nu}-E_1}$.  
Middle panels: The function $F(J)$ for the two possible transitions respectively. The different curves (solid, dashed and dotted) are for ${N=250,500,1000}$, \rmrkn{while ${NU=1}$.} 
Right panels: Dependence of~$T$ on the number of particles~$N$. 
It is determined by the height (solid line) of $F(J)$ 
for a non-optimized sweep, and by the area (dash line) of $F(J)$ 
for an optimized sweep, as implied by \Eq{eTdef}.
In the latter case the time of the sweep is $T/\lambda$, 
where $\lambda \approx 0.1$ is the required (scaled) rate.
}  
\label{fig:Sx}  
\end{figure*}

%%%%%%%%%%%%%%%%%%%%%%%%%%%%%%%%%%%%%%%%%%%%%%%%%%%%%%%%%%%%%%%%%%%%%%%%%%%
\section{The optimized adiabatic process}
\label{sec:Optimized}

\rmrk{We are dealing with an adiabatic scheme, and therefore the crucial question is whether the required time for its implementation is realistic. The purpose of this section is to provide an estimate for the required time, and to see how it scales with $N$. Roughly speaking the variation of the control parameter should be slow if the occupied adiabatic levels are getting close, and can be faster if they are weakly coupled or well spaced apart.}    

The upper panel of \Fig{fOutcome} shows the non-adiabatic leakage to higher levels. The gross dependence of the expectation value $\braket{n_{\text{ex}}}$ on the sweep rate is similar for the semiclassical and quantum calculations. 
\rmrk{\App{sec:Crossover} provides further details regrading the crossover from the adiabatic to the diabatic scenario.}
Our interest is in the quantum adiabatic regime, where we want to maintain ${\braket{n_{\text{ex}}} \ll 1}$. As can be seen in the figure, for a given bound that we want to impose on $\braket{n_{\text{ex}}}$, the maximum allowed sweep rate depends on $N$.

The quality of the phase estimation depends on the adiabaticity, and hence on our ability to allocate sufficient time for the splitting ($\Delta=0$) and branching ($\Delta \neq 0$) sweeps, during which $J$ is decreased to zero. On the other hand, practical considerations give preference to shorter sequences. This motivates us to suggest an optimized adiabatic process, where the sweep rate is continuously modified. Similar idea has been considered in \cite{CL2}. 

%%%%%%%%%%%%%%%%%%%%%%%%%%%%%%%%%%% 
\sect{Optimized splitting process}
The parameter that controls non-adiabatic transitions is {\em the scaled sweep rate:} 
\beq \label{eHdot}
\lambda(t) \ \ = \ \ \frac{\left|\bra{\mu}\dot{\mathcal{H}}\ket{\nu}\right|} {\left({E_{\mu}-E_{\nu}}\right)^2}   \, \, .
\eeq   
For the splitting process ($\Delta=0$), the parameter $\lambda$ is inspected as an indicator for undesired transitions between the ground state level ${\nu=1}$, 
and the excited level ${\mu=3}$, which has the same (even) parity. 
The adiabatic condition, ${\lambda(t) \ll 1}$, should be satisfied at any moment.
For an optimized protocol we set  ${\lambda(t) = \lambda = \const}$.   
Our protocol involves variation of $J(t)$, and therefore $\dot{\mathcal{H}} = -\dot{J}S_x$. 
Accordingly, for the optimized sweep we use  
\beq \label{eq:dotJ}
\dot{J} \ \ = \ \ \lambda  \left[ \frac{\left|\bra{3}S_x\ket{1}\right|}{\left({E_3-E_1}\right)^2} \right]^{-1}  
\equiv \ \  \frac{\lambda}{F(J(t))} \ \ .
\eeq   
The total time of an optimized sweep is 
\beq \label{eTdef}
\text{time} = \int_0^{J_{\text{max}}} \frac{dJ}{\dot{J}} 
 = \frac{1}{\lambda}  \int_0^{J_{\text{max}}} F(J) dJ   \ \equiv \ \frac{1}{\lambda} T \ .
\eeq

The lower panel of \Fig{fOutcome} shows the dependence of $\braket{n_{\text{ex}}}$ on the scaled sweep rate, $\lambda$. This dependence looks independent of $N$. This implies that in order to maintain adiabaticity the duration of the sweep process is required to be proportional to~$N$.
\Fig{fOutcome} also indicates that ${\lambda=0.1}$ is a good choice for the required adiabaticity. 

The only dimensional parameter in the calculation of $T$ is $U$, and therefore we expect 
\beq \label{eTnmr}
T \ \ = \ \ C\frac{N^{\alpha+1}}{NU} \ \ . 
\eeq
where $C$ is a numerical constant.  
Representative plots of $F(J)$, and the dependence of $\max{F(J)}$ and $T$ of $N$ are provided in \Fig{fig:Sx}.  
Recall that we fix the time units such that ${NU=1}$. Based on these results we conclude that ${\alpha \sim 0}$. 
Thanks to the optimization, the required time for an adiabatic sweep is shortened, albeit with the same scaling with respect to $N$, 
and for the numerical prefactor we get ${C \approx 0.07}$.
We note that the semiclassical analysis suggests ${N \ln(N)}$ dependence, but we cannot resolve the logarithmic correction numerically.

%%%%%%%%%%%%%%%%%%%%%%%%%%%%%%%%%%%%%%%%%%
\sect{Optimized branching process}
For the branching process ($\Delta \neq 0$) we can use the same optimization procedure. The bias is required to be small ${\Delta < U}$, such that EvenCat and OddCat states are adiabatically connected to $|N,0\rangle$ and $|0,N\rangle$, respectively, rather then ending at $|N,0\rangle$ and $|N{-}1,1\rangle$.  
Unlike the splitting process, here transitions from $E_1$ to $E_2$ have non-zero probability. In fact, we find that these transitions are more problematic than transition to $E_3$, as shown in the second row of panels in \Fig{fig:Sx}. For an optimized protocol we find that \Eq{eTnmr} applies with ${\alpha \sim 0}$ and ${C \approx 0.02}$. 

\begin{figure}[!b]
\includegraphics[width=8cm]{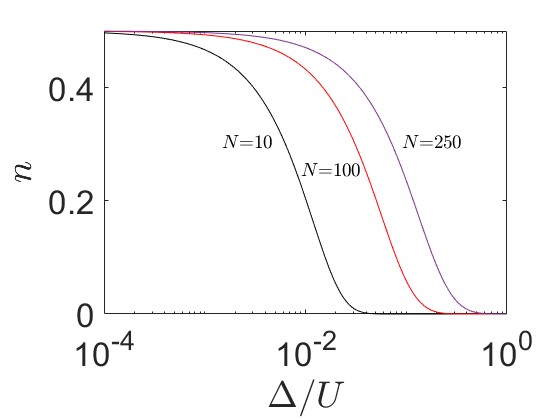}
\caption{{\bf The sensitivity of the splitting and branching processes to an energy bias.} 
The mean occupation $\braket{n}$ at the end of the sweep of $J$ down to zero is plotted versus the applied bias $\Delta$, with $\braket{n}=N/2$ indicating a cat state. The black, red, and purple curves are for ${N=10,100,250}$, respectively. Importantly, the sensitivity of generating a cat state using the adiabatic splitting method decreases with increased number of atoms.
} 
\label{fDeltaToll}  
\end{figure}

Whether we get splitting or branching depends on $\Delta/U$. It is important to figure out what is the tolerance of each of the processes to small deviations in setting $\Delta/U$. This determines the feasibility of the suggested protocol. The answer is provided in \Fig{fDeltaToll}, where we plot the mean occupation $\braket{n}$ at the end of the sweep of $J$ down to zero. This quantity can be used as an indicator to whether we get a cat-state with $\braket{n}=N/2$ or branching to either $\braket{n}=0$ or $\braket{n}=N$. We observe that for increased $N$, not only the tolerance does not deteriorate, but it even improves. Thus, in practice, one needs to tune the bias $\Delta$ to zero with an accuracy better than $0.01 \times U$.

%%%%%%%%%%%%%%%%%%%%%%%%%%%%%%%%%%%%%%%%%%
\sect{The merging process}
The protocol of \Fig{fProtocol} includes a merging stage that  constitutes a reversed version of the splitting. \rmrk{It maps the EvenCat and OddCat states to the ground state $\ket{X}$ and to its one-particle excitation-state respectively.}  We have verified (not displayed) that the same adiabatic condition applies.

%%%%%%%%%%%%%%%%%%%%%%%%%%%%%%%%%%%%%%%%%%
\sect{Turning on the bias}
The protocol of \Fig{fProtocol} includes a stage during which $\Delta$ is increased from zero to a finite value. Here non-adiabatic transitions due to large matrix element in \Eq{eHdot} are related to $\dot{\mathcal{H}} = -\dot{\Delta}S_z$. For large enough $J{=}J_\Delta$, the adiabatic states are roughly $\ket{1}=\ket{S_x{=}N/2}$ and $\ket{2}=\ket{S_x{=}N/2-1}$, the coupling is $\bra{1}S_z\ket{2} = \sqrt{N}/2$, and the adiabatic condition becomes 
\beq
\dot{\Delta} \ \ll \ \frac{(J_\Delta-UN)J_\Delta}{\sqrt{N}} \ \ .
\eeq 
This condition is easy to satisfy, and we have verified (not displayed) that it indeed ensures an adiabatic switch-on of the bias.

%%%%%%%%%%%%%%%%%%%%%%%%%%
%%%%%%%%%%%%%%%%%%%%%%%%%%

%%%%%%%%%%%%%%%%%%%%%%%%%%%%%%%%%%%%%%%%%%%%%%%%%%%%
%%%%%%%%%%%%%%%%%%%%%%%%%%%%%%%%%%%%%%%%%%%%%%%%%%%%
\section{Feasibility of the protocol}
\label{sec:Feasibility}

We consider a single Gaussian beam generating an attractive  optical trap. Let us denote the trap depth $V_0$ and its waist by $\sigma$. The harmonic trapping frequency close to its minimum is 
${\Omega_r=\sqrt{4V_0 / (m \sigma^2)} }$ in the radial direction, and 
${\Omega_z=\sqrt{ 2V_0 / (m Z_r^2)} }$ in the axial direction, 
where $m$ is the mass of the atom, 
${ Z_r = \pi \sigma^2 / \lambda_0 }$ is the Rayleigh range, and $\lambda_0$ is the trap wavelength. We require that a a Bose-Einstein Condensate (BEC) will be formed in the ground state of this potential. In order to drive the splitting process, one would split the potential into two traps and move away their central position. This can be done by employing two traps that are merged on a beam splitter and stirred independently by galvo mirrors, or by generating two traps using an acousto-optical modulator. Either way, the tunneling frequency between the two potentials is of the same order or lower than the harmonic trapping frequency in each beam. Additionally, we want to ensure that the changes we induce during the protocol do not cause excitations to higher eigenstates. Therefore, we require $J \lesssim \Omega_r$, where we assume that the tunneling occurs predominantly in the radial direction.

There are two parameters of each of the trapping potentials: $V_0$ and $\sigma$. To estimate the number of atoms, we take atom density of $\rho=10^{14}\,\text{cm}^{-3}$, which is typical for a BEC of ultracold atoms. The interaction energy scale is given by $U=4\pi\hbar^2 \rho |a|$, where $a$ is the s-wave scattering length. The sweeps of J should be done on a scale of $NU$, and therefore, we require $NU \lesssim \Omega_r$. \rmrk{For the gain in using a NOON state relative to a classical state to be significant, we aim to have $N\sim10-100$ atoms.}  

%For $^{87}$Rb, the scattering length is on the order of $100 a_0$, with $a_0$ being the Bohr radius. This yields an interaction energy scale on the order of $2\pi\times 1$ kHz, which means that for $N\sim10-100$, we get that $J$ needs to be a few tens of kHz. This translates into a stringent requirement for a tightly confined trap. However, the size of the condensate in a tightly confined trap is very small, which together with a given density means a small number of atoms.

To \rmrk{fulfill all the} requirements, one needs to allow tunability of the interaction energy. This can be achieved by working near a magnetic Feshbach resonance. We thus suggest working with a condensate of $^{39}$K atoms at the state $\ket{f=1,m_f=1}$, which have a broad resonance near $B_0=403.4$G \cite{NJP2007,PRR2020}. Importantly, the scattering length can be reduced all the way to the point where it crosses zero, around $B_{ZC}=350$G \cite{PRL2007}. Since adiabaticity requires that the sweep times are much larger than $U^{-1}$, we need to choose $a$ such that the interaction is not too small.

As an example, let us fix $a=-5a_0$ and choose a waist of $\sigma=10\mu$m, which does not require an objective with particularly high numerical aperture. With these numbers, we need to set the trap depth to $V_0=1.28 \times 10^{-27}$ Joule, which amounts to approximately $93 \mu$K. Assuming a trapping laser with a wavelength of $\lambda_l=1064$nm, far off the atomic resonance, this potential depth translates into a modest laser power of around $P_0=100$mW \cite{Advances2000}.

By the choice of parameters, we are working in the weak-interactions limit of the BEC \cite{ebooks}, hence we shall use a Gaussian approximation for its shape. Thus, the number of atoms in the condensate is given by $N=\rho \pi^{3/2} w_r^2 w_z$, with ${w_{r/z} = \sqrt{ \hbar / (m \Omega_{r/z})}}$ being the ground state extent in each direction \cite{ebooks}. With our choice of parameters, we obtain $N\approx50$, and $U/\hbar\approx 2\pi\times 86$Hz, and $J/\hbar\approx2\pi \times 4300$Hz, and $\Omega_r\approx2\pi\times 4478$Hz. The hierarchy of energy scales satisfy the required conditions. \rmrk{For an attractive BEC to be stable against collapse, $N<N_c$ with $N_c\approx 0.57 {(w_r^2 w_z)^{1/3}}/{|a|}$, see \cite{clps}. For our choice of parameters $N_c\approx 973$, therefore a BEC with $50$ atoms is stable. A plausible approach to generate a BEC of 50 atoms in a small trap is to load the trap from a large BEC that acts as a reservoir \cite{PRL1998}.}

Based on the results in \Fig{fOutcome} and \Fig{fig:Sx}, the sweep time of $J$ should exceed $10\times 0.07 (U/h)^{-1}\sim 8$ ms. The most challenging aspect of the experiment is the requirement to stabilize the bias between the two potentials to within $0.01\times U$, necessitating a relative stability of $\sim 0.5\cdot 10^{-6}$ in the laser power. \rmrk{This level of stability requires a high dynamic range for monitoring the beam’s power, achievable by combining multiple detectors with varying sensitivities.  If we limit the shortest response time of the laser intensity locking circuit to $\tau=1$ ms, more than sufficient for executing our proposed protocol, the relative shot noise in the laser intensity is $\sqrt{{h c}/{(\lambda_l \tau P_0)}}\approx 4 \cdot 10^{-8}$, an order of magnitude below the required stability. Thus, we conclude that the necessary laser intensity stability is feasible.}

\begin{figure}

\includegraphics[width=7cm]{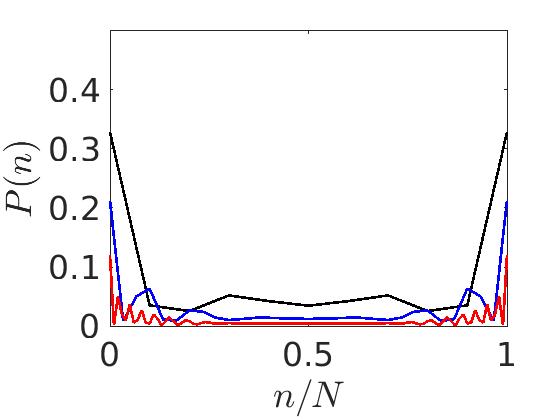} 

\vspace*{-47mm}

\includegraphics[width=3cm]{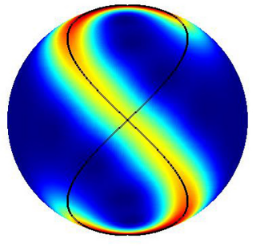} 

\vspace*{18mm}

\includegraphics[width=7cm]{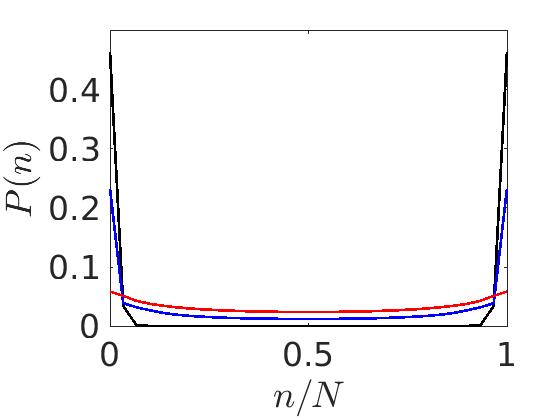} 

\vspace*{-47mm}

\includegraphics[width=3cm]{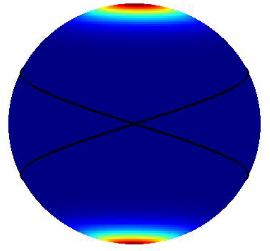} 

\vspace*{18mm}

\includegraphics[width=7cm]{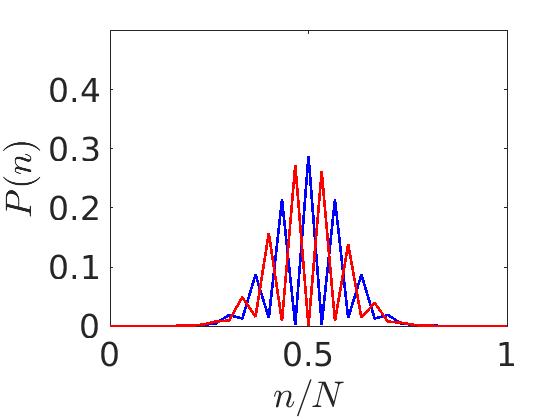} 

\vspace*{-47mm}

\hspace*{33mm}
\includegraphics[width=2cm]{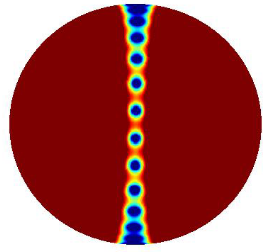} 

\vspace*{25mm}

\caption{{\bf Measured $n$ distribution.} 
{\em Upper panel:} The probability distribution $P(n)$ for the approximate NOON states that 
is prepared via non-adiabatic splitting of BEC, 
with ${N=10}$ (black), ${N=30}$ (blue), and ${N=100}$ (red) particles.
The inset is the Husimi fucntion of \Fig{fig:husimi} 
that represents the ${N=30}$ preparation.   
{\em Middle panel:}  The probability distribution $P(n)$ for the approximate NOON states that 
are prepared via adiabatic splitting of BEC with ${N=30}$ particles, 
and different sweep rates: 
${\dot{J}=1/6}$ (red), ${\dot{J}=1/8}$ (blue), and ${\dot{J}=1/100}$ (black).  
The inset is for the latter, the slowest sweep preparation.   
{\em Lower panel:} The probability distribution $P(n)$ for a $\pi/2$ rotated EvenCat state that has been 
produced by the slowest sweep of the middle panel. It includes a second curve (red) that is obtained if the accumulated phase is ${N\varphi=\pi}$ instead of ${N\varphi=0}$. 
The inset is of the same state of the middle panel, after $\pi/2$ rotation, 
and color scale that is streched by ${\times 10^7}$ in order to resolve the fringes   
(note that the Husimi function, by definition, is a smearead version 
of the Wigner funtion, therefore this color-scale streching in required).   
}
\label{fPn}  
\end{figure}
%%%%%%%%%%%%%%%%%%%%%%%%%%

%\newpage
%%%%%%%%%%%%%%%%%%%%%%%%%%%%%%%%%%%%%%%%%%%%%%%%%%%%
%%%%%%%%%%%%%%%%%%%%%%%%%%%%%%%%%%%%%%%%%%%%%%%%%%%%
\section{Discussion}
\label{sec:Discussion}

Several ways have been suggested how to prepare a NOON state. One way is to use the Schrodinger cat procedure, as in the experiment of Ref. \cite{opt1}, which requires the control and measurement of an internal degree of freedom. 
\rmrk{Other ways, that do not require an auxiliary degree of freedom,}  are e.g. {\em one-axis twisting}, as described after Eq(90) of Ref. \cite{RMV2018}. 
%
%Optional ways have been suggested for the preparation of an approximate NOON state, e.g. \cite{Fis1}. 
%
\rmrk{Optionally,} Ref.\cite{NonAdSplit} has suggested to stretch a Gaussian cloud along a separatrix, see \Fig{fig:husimi}b. The quality of the prepared cat state can be judged by looking at the final $P(n)$ distribution, see upper panel of \Fig{fPn}.  Note that for larger $N$ a deterioration is observed in the height of the $n{=}0$ and $n{=}N$ peaks. Ideally we would like to observe ${P(0)=P(N)=0.5}$.  
\rmrk{Irrespective of that, the above methods are sensitive to timing issues, and there are complications that are related to the lack of experimental control over the exact number of particles, and associated difficulties in parity measurements.}

{\em In view of the above, our preferred procedure for phase-estimation is an adiabatic version of non-linear splitting process}. Such method is robust, \rmrk{suitable for the tweezer interferometry setup}, not sensitive to timing issues, and provides better interferometry for large $N$. See illustration of the outcome in the middle panel of \Fig{fPn}b, where the $P(n)$ reflects an ideal NOON state.

At the second part of the protocol, the challenge is to extract the phase $\varphi$ of \Eq{ePrepC} from the measurement of quantum state $\psi(t)$.  There are several options how to achieve this goal. The conventional approach would be to use Ramsey $\pi/2$ rotation, as in standard interferometry. 
The $P(n)$ distribution will have strip interference pattern, as depicted in the lower panel of \Fig{fPn}. For an EvenCat, the central fringe is bright, while for a OddCat the central fringe is dark. While the difference between these outcomes is large, the problem lies in the fact that a single atom resolution in knowing the total number of atoms is required to distinguish the change of parity \cite{ParityMeas}. An experiment is presented in Ref. \cite{opt1}, where the number of fringes is ${N \sim 2}$. For higher $N$, this sensitivity to parity makes this approach extremely challenging for actual implementation.  

An optional way to perform phase estimation is to use the non-linear splitter in reverse. In such a procedure, the states $|+\rangle$ and $|-\rangle$ adiabatically become \rmrk{the original ground-state condensation and its single-particle excitation}. The difference between the outcome states is microscopic, and hence difficult to detect. 
The natural suggestion would be to use two-level-dynamics that maps  $|+\rangle$ and $|-\rangle$ to $|N,0\rangle$ and $|0,N\rangle$ respectively.  The latter are easily distinguished. The problem with this method is its slowness. Exponentially long time is required because the coupling between the two states is exponentially small in $N$, namely, ${J_{\text{eff}} = J e^{-cN}}$.   We note that there are also more exotic suggestions as in \cite{Fis2} that require to go beyond the two-orbital picture.    

{\em Our preferred procedure for phase-estimation is to reverse the splitting, and then to sweep down $J$ but with a non-zero bias.} 
To execute this process, one needs a good experimental control over $\Delta$ to allow ramping it to a small final non-zero value.   
\rmrk{This maps the  $\nu=1$ and  $\nu=2$ states} into the macroscopically distinct states $|N,0\rangle$ and $|0,N\rangle$ respectively. Importantly, the required time to reach the desired outcome scales as $N$ and not as $e^{cN}$. 

From the implementation perspective, a pertinent question emerges: {\em how can we distinguish between splitting, branching, and spontaneous symmetry breaking?} For example, in the experiment of \cite{QTPexp}, spontaneous symmetry breaking \cite{CL3} was reported and studied. As we have demonstrated, a minor uncontrolled imbalance $\Delta$ between the two wells during the splitting phase can indeed lead to spontaneous symmetry breaking. However, we have found that the required level of balance between the wells scales favorably with the number of particles. Our feasibility analysis has shown that with approximately 50 atoms, the imbalance should be kept below $10^{-6}$, which is challenging but achievable. Even when such a level is attained in the experiment, the only way to verify unambiguously coherent splitting is to perform the full interferometry scheme and observe $\varphi$-dependent output.

\ \\ \ \\ 

%%%%%%%%%%%%%%%%%%%%%%%%%%%%%%%%%%%%%%%%%%%%%%%%%%%%%%%
{\bf Acknowledgments} ---  
DC acknowledges support by the Israel Science Foundation, grant No.518/22. YS acknowledges support by the Israel Science Foundation (ISF), grant No. 3491/21, and by the Pazy Research Foundation.

\ \\ \ \\ 

%%%%%%%%%%%%%%%%%%%%%%%%%%%%%%%%%%%%%%%%%%%%%%%%%%%%%%%%%%%%%%%%%%%%%%%%%%%%%%%%%%%%%%%%%%%%%%%%%%%%%%%%%%%%%%%%
%\clearpage
\appendix

%%%%%%%%%%%%%%%%%%%%%%%%%%%%%%%%%%%%%%%%%%%%%%%%%%
\section{The characteristic frequencies}
\label{appA}

The phase space area $A(E)$ is define as 
\beq
A(E)  = \int_0^N dn \int_{-\pi}^{\pi}  d\phi  \Theta[E-\mathcal{H}(n,\phi)] 
\eeq
where $\Theta$ is the step function.
Here we use the conjugate coordinates $(n,\phi)$ relative to the X direction, meaning that ${n=0}$ indicates ${S_x=N/2}$. We use units such that formally ${\hbar=1}$, but in the WKB context this statement needs some clarification. The WKB quantization condition is 
\beq
A(E_{\nu}) \ = \ \left( \frac{1}{2}+\nu \right) h
\eeq      
where $h$ is the area of Planck cell. For the Bloch sphere it is implicit that the total area of phase space is $2\pi N = (N+1)h$. Form this follows that ${h=2\pi N/(N+1) \approx 2\pi }$. Note that the WKB quantization condition remains the same if we look on the complementary area, taking ${S_x=-N/2}$ as the origin.  The characteristic frequency is calculated as follows:  
\beq
\omega(E) \ = \ E_{\nu{+}1}{-}E_{\nu} \ \approx \ \left[\frac{A'(E)}{h}\right]^{-1}
\eeq
where the approximation ${h \approx 2\pi}$ is exact in the classical limit (large $N$). 

In our coordinates the Bloch vector is:
\beq
\vec{S} = \left( \frac{N}{2}{-}n , \sqrt{(N{-}n)n}\cos\phi , \sqrt{(N{-}n)n}\sin\phi \right) 
\ \ \ \ \ 
\eeq
and the Hamiltonian takes the form
\beq
\mathcal{H} = -J \left(\frac{N}{2}-n\right) - U(N{-}n)n \sin^2\phi
\eeq
The solution of $\mathcal{H} = E$ is 
\beq
&& n_{\pm}(E) \ = \ \frac{N}{2} -\frac{1}{2U \sin^2\phi} 
\nonumber \\
&& \times \left(J \mp \sqrt{J^2 + \left(4E + N^2 U \sin^2\phi \right)U \sin^2\phi }\right)
\eeq 
We take $E_x = -JN/2$ as the reference for the energy. 
This is the minimum energy for $J>J_c$ or the separatrix energy for $J<J_c$.    
Accordingly we substitute ${E \mapsto E_x + E }$, 
assume for simplicity ${E>E_x}$, 
and get the following integral expression 
\beq \label{eq:dA}
&& A'(E) = \ \ \frac{d}{dE} \left[ \int_{-\pi}^{\pi} n_{+}(E) d\phi \right]
\nonumber \\ 
&& \ \ \ \ = 4\int_{0}^{\pi/2} \!\!\! \frac{d\phi}{\sqrt{(J {-} NU\sin^2\phi)^2 + 4 E U \sin^2\phi }} 
\ \ \ 
\eeq 
For $J>J_c$ and setting $E=0$ we get as expected $A'(E)= 2\pi / \omega_J$, 
where $\omega_J$ is the frequency of the small oscillation at the minimum energy \Eq{eq:omegaJ}.  
At $J=J_c$ the integral diverges for ${E\rightarrow 0}$. 
and one can use the approximation 
\beq
A'(E) \ \approx \ \frac{4}{NU} \int_{0}^{\pi/2} \frac{d\phi}{\sqrt{\left(\phi-\frac{\pi}{2}\right)^4 + \frac{4E}{N^2U}}}  
\eeq
That leads to
\beq
A(E) \ \approx \  
 \frac{16\Gamma(\frac{1}{4})^2}{3\sqrt{2\pi}}
\left( \frac{E^3}{N^2U^3} \right)^{1/4}
\eeq
From the WKB quantization condition we get \Eq{eq:omegaC} for $\omega_c$.

For $J<J_c$ the integrand has a singular point at $\phi_r = \arcsin(J/(NU))$, 
and the $d\phi$ integration can be separated into 3 regions, namely
${[0,\phi_r{-}b] \cup [\phi_r{-}b,\phi_r{+}c] \cup [\phi_r{+}c,\pi/2]}$. 
In each region we use an appropriate approximation, namely,
\beq
&& 
\int_{0}^{\phi_r-b} \frac{d\phi}{J - NU\sin^2\phi} 
\nonumber \\ 
&& \ \ \ \  
= \frac{1}{2\omega_J} \ln \left( \frac{\sqrt{NU/J-1}+\cot(\phi_r-b)}{\cot(\phi_r-b)-\sqrt{NU/J-1}}\right)  
\nonumber \\ 
&& \ \ \ \ 
\approx \frac{1}{2\omega_J} \ln \left( \frac{2\omega_J}{NUb}\right)
\\ && 
\frac{1}{2\omega_J} \int_{\phi_r-b}^{\phi_r+c} \frac{d\phi}{\sqrt{(\phi-\phi_r)^2 + \frac{J E}{N\omega_J^2}}} 
\nonumber \\ 
&& \ \ \ \ 
\approx  
\frac{1}{2\omega_J} \ln \left(4bc \ \frac{N\omega_J^2}{J E} \right) 
\\ && 
\int_{\phi_r+c}^{\frac{\pi}{2}} \frac{d\phi}{J - NU\sin^2\phi} 
\nonumber \\ 
&& \ \ \ \ 
= \frac{1}{2\omega_J} \ln \left( \frac{\sqrt{NU/J-1}+\cot(\phi_r+c)}{\sqrt{NU/J-1}-\cot(\phi_r+c)}\right) 
\nonumber \\ 
&& \ \ \ \  
\approx \frac{1}{2\omega_J} \ln \left( \frac{2\omega_J}{NUc}\right)
\eeq
Adding the contributions together, the arbitrary borders $b$ and $c$ cancel out, 
and we get
\beq \label{eq:omegaE}
A'(E) \approx \frac{2}{\omega_J}\ln \left(\frac{16\omega_J^4}{NJU^2 \ E} \right) \ \ \equiv \ \ \frac{2\pi}{\omega(E)}
\eeq
The WKB estimate \Eq{eq:omegaX} for the frequency $\omega_x$ 
is obtained by regarding this expression as constant 
with ${E \mapsto \omega_J }$. This can be regarded 
as a leading order approximation of an iterative scheme.

%%%%%%%%%%%%%%%%%%%%%%%%%%%%%%%%%%%%%%%%%%%%%%%%%%%%%%%%%%
\section{Crossover to diabatic scenario}
\label{sec:Crossover}

In the semi-classical picture the cloud spreads along the separatrix. This is demonstrated in \Fig{fig:husimi}. 
The angular spread in this illustration is $\theta \sim \pi$. In the adiabatic limit we can be estimate $\theta$ analytically as follows:
\beq\label{eq:cdotJ}
\theta = \int \omega_x dt = \frac{1}{\dot{J}} \int_0^{J_c} \omega_x dJ \sim \frac{\pi^2}{8} \left[\frac{(NU)^2}{\ln(N)}\right] \frac{1}{\dot{J}} \ \ .
\eeq
In the above integral, the range of integration reflects a sweep process that starts at $J=J_c=NU$ and ends at $J=0$. The frequency $\omega_x$ depends on $J$ as implied by \Eq{eq:omegaX}. Consequently we get the final estimate for $\theta$.  
This allows the determination of the range where a crossover from adibatic to diabatic evolution is observed in \Fig{fOutcome}. Namely, we determine the $\dot{J}$ borders of the crossover region from the equation $\theta = 2\pi s$, with $s=0.1$ for the diabatic border and $s=0.9$ for the adiabatic border. 

In the adiabatic range, the leakage $\braket{n_{\text{ex}}}$ to higher levels is zero, because the final state is a superposition of the $S_z=-N/2$ and $S_z=N/2$ states for which $n_{\text{ex}}=0$. In the diabatic range, the sudden approximation implies that the state stays at  $S_x=N/2$, therefore  $n_{\text{ex}}\sim N/2$.

\newpage
%\clearpage
%%%%%%%%%%%%%%%%%%%%%%%%%%%%%%%%%%%%%%%%%%%%%%%%%%%%%%%%%%%%%%%%%%%%%%%%%%%%%%%%%%%%%%%%%%%%%%%%%%%%%%%%%%%%%%%%

%%%%%%%%%%%%%%%%%%%%%%%%%%%%%%%%%%%%%%%%%%%%%%%%%%%%%%%%%%%%%%%%
\end{document}